\documentclass[12pt]{article}
\usepackage[margin=1.25in]{geometry}
\usepackage[T1]{fontenc}
\usepackage[utf8]{inputenc}
\usepackage{amsmath,graphicx}
\usepackage{gensymb}
\usepackage{tikz}
\usepackage[round]{natbib}
\usepackage{multicol,caption}
\renewcommand{\baselinestretch}{1.0}
\newcommand{\y}{{\bf y}}

\newcommand{\Y}{{\bf Y}}
\newenvironment{Figure}
  {\par\medskip\noindent\minipage{\linewidth}}
  {\endminipage\par\medskip}
\usepackage{fancyvrb}
\usepackage{float}
\usepackage{setspace}
\usepackage{etoolbox}
\usepackage{bm}
\usepackage{comment}
\usepackage{subfig}
\title{Bayesian Estimation of Attribute Disclosure Risks in Synthetic Data with the \texttt{AttributeRiskCalculation} R Package}
\author{Ryan Hornby\footnote{Vassar College, Box 2785, 124 Raymond Ave, Poughkeepsie, NY 12604, United States, rhornby@vassar.edu} and Jingchen Hu\footnote{Vassar College, Box 27, 124 Raymond Ave, Poughkeepsie, NY 12604, United States, jihu@vassar.edu}}

\begin{document}
\BeforeBeginEnvironment{Verbatim}{\def\baselinestretch{1}}
\maketitle

\begin{abstract}
    Synthetic data is a promising approach to privacy protection in many contexts. A Bayesian synthesis model, also known as a synthesizer, simulates synthetic values of sensitive variables from their posterior predictive distributions. The resulting synthetic data can then be released in place of the confidential data. An important evaluation prior to synthetic data release is its level of privacy protection, which is often in the form of disclosure risks evaluation. Attribute disclosure, referring to an intruder correctly inferring the confidential values of synthetic records, is one type of disclosure that is challenging to be computationally evaluated. In this paper, we review and discuss in detail some Bayesian estimation approaches to attribute disclosure risks evaluation, with examples of commonly-used Bayesian synthesizers. We create the \texttt{AttributeRiskCalculation} R package to facilitate its implementation, and demonstrate its functionality with examples of evaluating attribute disclosure risks in synthetic samples of the Consumer Expenditure Surveys.
\end{abstract}

{\bf{keywords}}: Attribute disclosure risks, Bayesian synthesizer, Importance sampling, Privacy protection, Synthetic data

{\color{red} 

}

\section{Introduction}
\label{intro}

Respondent-level data, also known as microdata, are extremely important to many disciplines including social and behavioral sciences. However, these microdata sometimes contain 
sensitive information about records in the dataset. Therefore, including such information in its original form in a released dataset could lead to identification of an individual and / or disclosure of private information. To avoid such privacy and confidentiality compromises while maintaining the usefulness of released data, data disseminators could opt to create synthetic dataset(s) based on models of the confidential data, where the sensitive information is replaced with synthetic values (\citet{Rubin1993synthetic, Little1993synthetic, ReiterRaghu2007}). Depending on the protection goals, data disseminators can choose between two flavors of synthetic data: if a subset of the variables is deemed sensitive, partially synthetic data containing synthetic values of these sensitive variables can be created and released \citep{Little1993synthetic}; if all variables are deemed sensitive, then fully synthetic data with every variable being synthesized can be created release \citep{Rubin1993synthetic}. See \citet{Drechsler2011book} for a detailed introduction to synthetic data for privacy protection. 

Once synthetic data are generated, data disseminators perform two types of evaluation before their release. The first type is the utility evaluation of the synthetic data, i.e. how useful the synthetic data is for users. There are global utility measures, which focus on measuring the distance between the confidential data and the synthetic data \citep{Woo2009JPC, Snoke2018JRSSA}. There are also analysis-specific utility measures, which rely on specific analyses users might conduct and compare results obtained from the confidential data with those from the synthetic data \citep{KarrKohnenOganianReiterSanil2006}. Overall, the synthetic data research community has done extensive research on utility evaluation and several methods are widely-used, including the propensity score global utility of \citet{Woo2009JPC} and interval overlap analysis-specific utility of \citet{KarrKohnenOganianReiterSanil2006}. 

The second type of evaluation is the level of privacy protection offered by the synthetic data, which is often in the form of disclosure risks evaluation. There exist two common disclosures in synthetic data: (a) identification disclosure, where intruder correctly identifies records of interest in the synthetic data using additional information through external databases; and (b) attribute disclosure, where intruder correctly infers the confidential values of synthetic records given the synthetic data. It is generally believed that only the attribute disclosure exists in fully synthetic data, while both types of disclosures are present in partially synthetic data \citep{Hu2019TDP}.

For identification disclosure risks evaluation of partially synthetic data, \citet{ReiterMitra2009} proposed estimation methods based on matching with available information from external databases. \citet{HornbyHu2020IR} recently reviewed the matching-based methods and extended them to multivariate synthetic data. The authors also created the \texttt{IdentificationRiskCalculation} R package for implementation \citep{IdentificationRiskCalculation}. 

For attribute disclosure risks evaluation, \citet{HuReiterWang2014PSD} laid out a Bayesian framework for its estimation. For a record of interest, consider the most conservative scenario that the intruder knows all records but this particular record of interest. Next, the data disseminators create a collection of guesses for this record's confidential value(s), which includes the true confidential value(s). Data disseminators then proceed to estimate the posterior probability of each guess, conditional on the simulated and released synthetic data, all other records available to the intruder, and any other information. In this way, data disseminators are able to evaluate the posterior probability of guessing the true confidential value(s) correctly, compared to other possible guesses. The estimation process also utilizes the importance sampling technique to avoid fitting the chosen synthesis model on each possible dataset (containing the confidential values of all other records and the guess being evaluated for the record of interest) when approximating the likelihood in the evaluation process. \citet{HuReiterWang2014PSD} proposed the framework and presented the estimation detail with a particular synthesizer, the Dirichlet Process mixtures of products of multinomials. Other works using a similar framework also described the estimation detail with particular synthesizers \citep{Reiter2014framework, Paiva2014SM}. 

While this approach to attribute disclosure risks evaluation is applicable to any Bayesian synthesizer, its computational details could be challenging to implement. In this work, we wish to review the general framework in a gentle and approachable manner, while provide illustrations with sample R scripts to several commonly-used Bayesian synthesizers, including a multinomial logistic regression synthesizer for categorical variables, a linear regression synthesizer for continuous variables, and a Poisson regression synthesizer for count variables. Moreover, we create the \texttt{AttributeRiskCalculation} R package for the implementation of these Bayesian estimation methods for attribute disclosure risks evaluation, if users choose these routines instead of writing evaluation scripts by themselves \citep{AttributeRiskCalculation}. We demonstrate how to use these routines from the \texttt{AttributeRiskCalculation} R package with applications to a Consumer Expenditure Surveys (CE) sample, where multiple variables of different types are present.



Section \ref{intro:seq} provides a succinct review of the sequential synthesis approach, where multiple variables are deemed sensitive in a dataset and therefore to be synthesized. We take this commonly-used approach in synthesizing more than one variables in some of our illustrative examples. Moreover, the details of sequential synthesis are important to construct the Bayesian estimation methods for attribute disclosure risks evaluation, which are reviewed and discussed in this work.

\subsection{Sequential synthesis}
\label{intro:seq}


Sequential synthesis is a common technique to create synthetic datasets with more than one synthetic variable (e.g. \citet{kinney_reiter_reznek_miranda_jarmin_abowd_2011}). The general idea is that data disseminators can fit a series of conditional distributions of one sensitive variable at a time. Without loss of generality, assume there are three variables, $(\y_1, \y_2, \y_3)$ in a confidential dataset. Among them, $\y_2$ and $\y_3$ are deemed sensitive and to be synthesized, while $\y_1$ is considered insensitive and un-synthesized. Assume the synthesis order is to synthesize $\y_2$ given $\y_1$ first and then $\y_3$ given $(\y_2, \y_1)$. The joint distribution of all variables can then be expressed in the product of the following series of conditional distributions:
\begin{equation}
    f(\y_1, \y_2, \y_3) = f(\y_1) f(\y_2 \mid \y_1) f(\y_3 \mid \y_1, \y_2).
\end{equation}

To start the sequential synthesis process, first choose a suitable Bayesian synthesizer for $\y_2 \mid \y_1$ (i.e. a synthesizer for $\y_2$ where $\y_1$ is used as a predictor) and fit it on the confidential data $(\y_1, \y_2)$. Simulate synthetic values of $\y_2$ from its posterior predictive distribution using confidential $\y_1$, denoted as $\tilde{\y}_2$. Next, choose a suitable Bayesian synthesizer for $\y_3 \mid \y_1, \y_2$ and fit on the \emph{confidential} data $(\y_1, \y_2, \y_3)$. Simulate synthetic values of $\y_3$ from its posterior predictive distribution using confidential $\y_1$ and \emph{synthetic} $\tilde{\y}_2$. The resulting $(\y_1, \tilde{\y}_2, \tilde{\y}_3)$ is one synthetic dataset. 


The remainder of the paper is organized as follows. In Section \ref{methods}, we describe the general methods to calculate attribute disclosure risks and then illustrate three Bayesian synthesizers with sample R scripts. We also review the use of importance sampling. Section \ref{package} gives an overview of the \texttt{AttributeRiskCalculation} R package. We then demonstrate the routines in the package with applications to calculating attribute disclosure risks of simulated synthetic data for a CE sample in Section \ref{app}. We end with Section \ref{conclusion} with a few concluding remarks. Sample scripts of using either the \texttt{rstanarm} \citep{rstanarm} or the \texttt{brms} \citep{brms} R packages to fit several Bayesian synthesizers are included in the Appendix for interested readers.


\section{Calculation methods and computational details}
\label{methods}

We describe the general approach to attribute disclosure risks calculation in Section \ref{methods:AR_cal}. We then illustrate the computational details for three Bayesian synthesizers in Section \ref{methods:computing}. In Section \ref{methods:importance}, we show the computational details for the importance sampling technique of the Bayesian linear regression synthesizer presented in Section \ref{methods:computing:2cont} for illustration.

\subsection{General approach to attribute disclosure risks calculation}
\label{methods:AR_cal}

Our description of the general approach is similar to that in \cite{HuReiterWang2014PSD}. Denote our confidential dataset as $\y$, consisting of $n$ individuals with $p$ variables. These variables will either be synthesized ($\y^s$) or un-synthesized ($\y^{us}$) in the released dataset(s) $\tilde{\Y}$. Note that $\y^{us}$ could be an empty set of variables in the case of fully synthetic data. We aim to calculate the probability of guessing the true confidential value(s) of some synthesized variable(s) for individual of interest, $i$, given synthetic data $\tilde{\Y}$ and any other auxiliary information the intruder may know. We divide the auxiliary information into knowledge of individuals in the dataset (denoted as $A$) and knowledge of the process of generating synthetic $\tilde{\Y}$ from confidential $\y$ (denoted as $S$). Let the set of knowledge the intruder has about $\y$ be $K= \{\y^{us}, A, S\}$. Therefore, the intruder aims to calculate the following probability:
\begin{align}
    p({\bf Y}_i^s = \y^* \mid \tilde{\Y}, K),
    \label{eq:dist}
\end{align}
where $\Y_i^s$ is the random variable that represents the intruder's uncertain knowledge of the true confidential value(s) $\y_i^s$, and $\y^*$ is the intruder's guess for the true confidential value(s). Without loss of generality, we use $\mathbf{y}_i^s$ to represent the vector of synthesized variables, for the general case of multiple sensitive variables being synthesized. When only one variable is synthesized, $\mathbf{y}_i^s$ reduces to $y_i^s$. Using Bayes' rule, we rewrite Equation (\ref{eq:dist}) as:
\begin{align}
    p({\bf Y}_i^s = \y^* \mid \tilde{\Y}, K) \propto p(\tilde{\Y} \mid \Y_i^s = \y^*, K) p(\Y_i^s = \y^* \mid K),
    \label{eq:bayes}
\end{align}
where $p(\tilde{\Y} \mid \Y_i^s = \y^*,K)$ is the probability of simulating synthetic $\tilde{\Y}$ given guess $\Y_i^s = \y^*$ and intruder's knowledge $K$. $p(\Y_i^s = \y^* \mid K)$ is the chosen prior distribution for guess $\Y_i^s = \y^*$. In our presentation, we assume a uniform prior for $p(\Y_i^s = \y^* \mid K)$, which means that estimating the posterior probability of $p({\bf Y}_i^s = \y^* \mid \tilde{\Y}, K)$ becomes estimating the likelihood portion, $p(\tilde{\Y} \mid \Y_i^s = \y^*, K)$. 

For $K$, we note that information about $S$, the synthesis process, can sometimes be publicly available, such as the synthesis models published in \citet{kinney_reiter_reznek_miranda_jarmin_abowd_2011}. As a worst case scenario, we will assume the intruder has extensive knowledge of the synthesis methods. Moreover, as a worst case scenario, we will assume the intruder knows the confidential values of synthetic $\y^{s}$ for all but record $i$. We denote this worst case knowledge be $K^w$. 

Equation (\ref{eq:bayes}) with our worst case scenario of $K^w$ is $p(\tilde{\Y} \mid \Y_i^s = \y^*, K^w)$. 
For notation simplicity, we work with one dataset $\tilde{\Y}$ and note that $p(\tilde{\Y} \mid \Y_i^s = \y^*, K^w) = \prod_{l=1}^m p(\tilde{\Y}_l \mid\Y_i^s = \y^*, K^w)$ if $m > 1$ synthetic datasets are simulated. To estimate $p(\tilde{\Y} \mid \Y_i^s=\y^*,K^w)$, we use posterior parameter draws from fitting the synthesizer on the confidential data. Denote the collection of the model parameters as $\Theta$:
\begin{align}
    p(\tilde{\Y} \mid \Y_i^s = \y^*, K^w) = \int p(\tilde{\Y} \mid \Y_i^s = \y^*, K^w, \Theta) p(\Theta \mid \Y_i^s = \y^*, K^w) d\Theta.
    \label{eq:MonteCarlo}
\end{align}
Typically a Monte Carlo approximation would be used for estimating Equation (\ref{eq:MonteCarlo}). However, doing so requires fitting the Bayesian synthesizer on $(\Y_i^s = \y^*, K^w)$ for every guess $\Y_i^s = \y^*$. That is, the Bayesian synthesizer needs to be estimated on $(\y^*, \y_{-i})$, a dataset consisting of the guess $\y^*$ for $\Y_i^s$ for record $i$ and confidential values for all other records, denoted by $\y_{-i}$. When multiple guesses are present for each record and multiple records need to be evaluated, this Monte Carlo approximation approach could be computationally expensive. 

To tackle this computation challenge, we apply the importance sampling technique for the estimation of $g(\Theta) = p(\tilde{\Y} \mid \Y_i^s = \y^*, K^w)$ of Equation (\ref{eq:MonteCarlo}). Specifically,
\begin{align}
    E[g(\Theta)] \approx \frac{1}{H}\sum_{h=1}^H g(\Theta^{(h)}) \frac{f(\Theta^{(h)})/f^*(\Theta^{(h)})}{\sum_{k=1}^H f(\Theta^{(k)})/f^*(\Theta^{(k)})},\label{eq:sampling}
\end{align}
where our distribution of interest is $g(\Theta)=p(\tilde{\Y} \mid \Y_i^s = \y^*, K^w)$ and for the $h$th posterior parameter draws $\Theta^{(h)}$, $g(\Theta^{(h)}) = p(\tilde{\Y} \mid \Y_i^s = \y^*, K^w, \Theta^{(h)})$, where $H$ is the number of posterior sample draws of $\Theta$. Moreover, $f(\Theta^{(h)}) = p(\Theta^{(h)} \mid \Y_i^s = \y^*, K^w)$ and $f^*(\Theta^{(h)}) = p(\Theta^{(h)} \mid \y, S)$. 

Recall that our density of interest for $\Theta$ is $f(\Theta) = p(\Theta \mid \Y_i^s = \y^*, K^w)$. The importance sampling technique utilizes a convenient distribution, $f^*(\cdot)$, that is readily available and differs slightly from $f(\cdot)$. For us, this convenient distribution is $f^*(\Theta) = p(\Theta \mid \y, S)$, the posterior distribution of the parameters $\Theta$ after fitting the synthesizer on the confidential data $\y$. These posterior parameter draws are available from the synthesis process, and we can use them in the importance sampling step in Equation (\ref{eq:sampling}) for approximation of $p(\tilde{\Y} \mid \Y_i^s = \y^*, K^w)$ in a computationally efficient manner. The choice of $H$ needs to be large enough for a good approximation, but not too large and creates computational burden.

Next, we present the computational details for three commonly-used Bayesian synthesizers. In particular, we break down the process and show how to calculate $g(\Theta^{(h)})$, $f(\Theta^{(h)})$, and $f^*(\Theta^{(h)})$ for each Bayesian synthesizer. 

\subsection{Computational details for three Bayesian synthesizers}
\label{methods:computing}

We go through three examples to illustrate the computational details of calculating $g(\Theta^{(h)})$, $f(\Theta^{(h)})$, and $f^*(\Theta^{(h)})$ for three commonly-used Bayesian synthesizers: a multinomial logistic regression synthesizer for categorical variables (Section \ref{methods:computing:1cat}), a linear regression synthesizer for continuous variables (Section \ref{methods:computing:2cont}), and a Poisson regression synthesizer for count variables (Section \ref{methods:computing:1count1bin1cont}).


\subsubsection{One synthetic categorical variable}
\label{methods:computing:1cat}


Suppose we are planning to release a single synthetic dataset of $n$ records, where we have synthesized one categorical variable $\y$ of $k$ levels. Further suppose that we synthesize this variable from a multinomial logistic regression synthesizer with a predictor $\y_p$. After Markov chain Monte Carlo (MCMC) estimation of the synthesizer on the confidential data, we obtain posterior parameter draws of $\Theta$. Synthetic values of $\y$ can be simulated from its posterior predictive distribution given draws of $\Theta$, resulting in a partially synthetic dataset of $(\tilde{\y}, \y_p)$. For our presentation, we consider $\y_p$ as binary or continuous for notation simplicity. 

To calculate attribute disclosure risks for the $i$th record, we first create a collection of guesses $\{y_1^*, \cdots, y_G^*\}$, where $G$ is the number of guesses (for a synthetic categorical variable, each guess is a scalar). We ensure that the true confidential value for record $i$, $y_i$, is in this collection. For a categorical variable, it is reasonable to enumerate all of its possible levels as the collection, i.e. $G = k$, a strategy we take in our application to a CE sample in Section \ref{app:oneCat}.

To estimate the posterior probability of each guess in collection $\{y_1^*, \cdots, y_G^*\}$, we calculate the quantities of $g(\Theta^{(h)})$, $f(\Theta^{(h)})$, and $f^*(\Theta^{(h)})$ to be used in the importance sampling step for the $h$th posterior draw of $\Theta$. Here our $g(\Theta^{(h)})$ is:
\begin{gather*}
    g(\Theta^{(h)}) = \prod_{i=1}^n M(\Tilde{y}_i, {\bf p}_i^{(h)}),
\end{gather*}
where $\tilde{y}_i$ is synthetic value for record $i$, $M(\cdot)$ is the probability mass function (pmf) of a multinomial distribution, and ${\bf p}_i^{(h)}$ is the vector of probabilities for each level of $y_i$, calculated from $\Theta^{(h)}$. We can calculate this quantity with the following R code:
\begin{Verbatim}[frame=single]
g_h = prod(p[h, y_syn])
\end{Verbatim}
Here \texttt{p} is a $n\times (k-1)$ matrix, where $k$ is the number of levels, and \texttt{y\_syn} is the synthetic vector of length $n$. Note that since we are computing the density of one record at a time, we can obtain the probability associated with that category instead of using a call to \texttt{dmultinom()}.

Next, our $f(\Theta^{(h)})$ and $f^*(\Theta^{(h)})$ functions are:
\begin{align*}
    f(\Theta^{(h)}) &= M(y_g^*, {\bf p}_i^{(h)}),\\
    f^*(\Theta^{(h)}) &= M(y_i, {\bf p}_i^{(h)}),
\end{align*}
where $y_g^*$ is the guess being evaluated and $y_i$ is the true confidential value of record $i$. 
These can be calculated with the following R code:
\begin{Verbatim}[frame=single]
f_h = p[h, y_guess]
f_hs = p[h, y_i]
\end{Verbatim}
Here \texttt{y\_guess} is the current guess being evaluated and \texttt{y\_i} is the true value of our variable for record $i$. 

\subsubsection{Two synthetic continuous variables (sequential synthesis)}
\label{methods:computing:2cont}

Suppose we want to synthesize two continuous variables $\y_1$ and $\y_2$ of $n$ records. According to the sequential synthesis strategy reviewed in Section \ref{intro:seq}, we first synthesize $\y_1$ with a linear regression synthesizer with no predictors, and next synthesize $\y_2$ with a linear regression synthesizer using $\y_1$ as a predictor. The sequential synthesis process creates a fully synthetic dataset of $(\tilde{\y}_1, \tilde{\y}_2)$. 

To calculate the attribute disclosure risks for the $i$th record, we start with creating a collection of our guesses $\{y_{11}^*, \cdots, y_{1G_1}^*, y_{21}^*, \cdots, y_{2G_2}^*\}$, where $G_1$ and $G_2$ are the number of guesses for the confidential values of each variable, respectively. Therefore, there are $G_1 \times G_2$ number of guesses for the confidential values of the pair of synthetic $(\tilde{y}_{i1}, \tilde{y}_{i2})$. As before, we ensure that the true confidential value pair, $(y_{i1}, y_{i2})$, is in this collection of guesses. Unlike categorical variables where different levels are natural candidates for these guesses, for continuous variables, there are in theory infinite number of guesses available. We recommend creating the collection of guesses in the neighborhood of the true confidential value with a reasonable number of guesses. For example for $y_{i1}$, we can create a neighborhood interval of $[y_{i1} \times 0.9, y_{i1} \times 1.1]$ (i.e. within a 20\% radius of $y_{i1}$) and then select $G_1$ equally-spaced guesses from this interval. Similar approach can be applied to $y_{i2}$, resulting $G_1 \times G_2$ neighboring pairs for $(y_{i1}, y_{i2})$ as our guesses. We adopt this strategy in our CE application in Section \ref{app:twoCont}, with $G_1 = G_2$.

Now we proceed with quantities for importance sampling step. First, $g(\Theta^{(h)})$ is:
\begin{align*}
    g(\Theta^{(h)}) = \prod_{i=1}^n \left(\phi(\Tilde{y}_{i1},\mu^{(h)}, \sigma^{(h)}_1)\phi(\Tilde{y}_{i2}, \beta_{0}^{(h)} + \Tilde{y}_{1_i}\beta_{1}^{(h)}, \sigma^{(h)}_2)\right),
\end{align*}
where $\tilde{y}_{i1}$ and $\tilde{y}_{i2}$ are the synthesized values of record $i$, $\Theta^{(h)} = (\mu^{(h)}, \sigma_1^{(h)}, \beta_{0}^{(h)}, \beta_{1}^{(h)}, \sigma_2^{(h)})$, and $\phi(\cdot)$ is the probability density function (pdf) of a normal distribution. $g(\Theta^{(h)})$ can be calculated with the following R code:
\begin{Verbatim}[frame=single]
g_h = prod(dnorm(y_1_syn, mu[h], sigma_1[h]) *
      dnorm(y_2_syn, b0[h] + y_1_syn[i] * b1[h], sigma_2[h]))
\end{Verbatim}
Here \texttt{dnorm()} is the normal density function, \texttt{y\_1\_syn} and \texttt{y\_2\_syn} are synthetic vectors for $\y_1$ and $\y_2$ of length $n$ respectively, and \texttt{b0}, \texttt{b1}, \texttt{mu}, \texttt{sigma\_1}, and \texttt{sigma\_2} are our parameter draws, all of length $H$. 

Next, our $f(\Theta^{(h)})$ and $f^*(\Theta^{(h)})$ functions are:
\begin{align*}
    f(\Theta^{(h)}) &= \phi(y_{1g}^*,\mu^{(h)}, \sigma^{(h)}_1)\phi(y_{2g}^*, \beta_{0}^{(h)} + y_{1g}^*\beta_{1}^{(h)}, \sigma^{(h)}_2),\\
    f^*(\Theta^{(h)}) &= \phi(y_{i1},\mu^{(h)}, \sigma^{(h)}_1)\phi(y_{i2}, \beta_{0}^{(h)} + y_{1_i}\beta_{1}^{(h)}, \sigma^{(h)}_2),
\end{align*}
where $y^*_{1g}$ and $y^*_{2g}$ are the pair of guesses being evaluated and $y_{i1}$ and $y_{i2}$ are the true confidential values for record $i$. The following R code calculates these quantities:
\begin{Verbatim}[frame=single]
f_h = dnorm(y_1_guess, mu[h], sigma_1[h]) *
      dnorm(y_2_guess, b0[h] + y_1_guess * b1[h], sigma_2[h]) 
f_hs = dnorm(y_1_i, mu[h], sigma_1[h]) *
       dnorm(y_2_i, b0[h] + y_1_i * b1[h], sigma_2[h])
\end{Verbatim}
Here \texttt{y\_1\_guess} and \texttt{y\_2\_guess} are the guesses currently being evaluated for record $i$ and \texttt{y\_1\_i} and \texttt{y\_2\_i} are its true confidential values.

\subsubsection{Synthetic count and continuous variables (sequential synthesis)}
\label{methods:computing:1count1bin1cont}


Our last example is creating synthetic values of a count variable $\y_{count}$ and a continuous variable $\y_{cont}$ of $n$ records. Our sequential synthesis strategy will first synthesize $\y_{cont}$ with a linear regression synthesizer with a predictor $\y_p$, and next synthesize $\y_{count}$ with a Poisson regression synthesizer using both $\y_{cont}$ and $\y_p$ as predictors. The sequential synthesis process creates a partially synthetic dataset of $(\y_p, \tilde{\y}_{cont}, \tilde{\y}_{count})$. As before for notation simplicity, we consider $\y_p$ as binary or continuous. 


As usual, to calculate the attribute disclosure risks for the $i$th record, we start with creating a collection of our guesses $\{y_{cont, 1}^*, \cdots, y_{cont, G_1}^*, y_{count, 1}^*, \cdots, y_{count, G_2}^*\}$, where $G_1$ and $G_2$ are the number of guesses for the confidential values of the continuous and count variables, respectively, resulting in $G_1 \times G_2$ number of guesses for the pair of synthetic $(\tilde{y}_{cont, i}, \tilde{y}_{count, i})$. The true confidential value pair, $(y_{cont, i}, y_{count, i})$, is included in this collection. We recommend the interval-based approach to select $G_1$ guesses for the continuous variable, discussed in Section \ref{methods:computing:2cont}. For the count variable, we could either exhaust all possible counts in the dataset (if it is not too large) or pick $G_2$ guesses from the neighborhood of $y_{count, i}$. Our CE application in Section \ref{app:1count1bin1cont} takes the first strategy since there are only 8 possible count values for that CE variable.


To proceed, we first have $g(\Theta^{(h)})$ as
\begin{eqnarray*}
    g(\Theta^{(h)}) = \prod_{i=1}^n \Bigg (&&\phi(\tilde{y}_{cont, i},\beta_{0, cont}^{(h)} + \beta_{1, cont}^{(h)}y_{p, i}, \sigma^{(h)}) \\
    &&P(\tilde{y}_{count, i}, link(\beta_{0, count}^{(h)} + \beta_{1, count}^{(h)}\tilde{y}_{count, i} + \beta_{2, count}^{(h)}y_{p, i})) \Bigg ),
\end{eqnarray*}
where $\phi(\cdot)$ is the normal pdf, $\Theta^{(h)} = (\beta_{0, count}^{(h)}, \beta_{1, count}^{(h)}, \sigma^{(h)}, \beta_{0, count}^{(h)}, \beta_{1, count}^{(h)}, \beta_{2, count}^{(h)})$, and $P(\cdot)$ is the Poisson pmf. Note that since the predictor variable is not synthesized, its confidential value $y_{p, i}$ is in these calculations. Using $link(\cdot) = exp(\cdot)$, the following R code calculates $g(\Theta^{(h)})$:

\begin{Verbatim}[frame=single]
g_h = prod(dnorm(y_cont_syn, b0_cont[h] + b1_cont[h] * y_p, sigma[h])
           * dpois(y_count_syn, 
                   exp(b0_count[h] + b1_count[h] * y_cont_syn
                       + b2_count[h] * y_p)))
\end{Verbatim}
Here \texttt{dpois()} is the Poisson mass function, \texttt{y\_cont\_syn} and \texttt{y\_count\_syn} are synthetic vectors for $\y_{cont}$ and $\y_{count}$ of length $n$ respectively, and \texttt{b0\_cont}, \texttt{b1\_cont}, \texttt{sigma}, \texttt{b0\_count}, \texttt{b1\_count}, and \texttt{b2\_count} are our parameter draws, all of length $H$.

Next, our $f(\Theta^{(h)})$ and $f^*(\Theta^{(h)})$ functions are:
\begin{align*}
    f(\Theta^{(h)}) = &\phi(y^*_{cont, i},\beta_{0, cont}^{(h)} + \beta_{1, cont}^{(h)}y_{p, i}, \sigma^{(h)})\\&P(y^*_{count, i}, link(\beta_{0, count}^{(h)} + \beta_{1, count}^{(h)}y^*_{count, i} + \beta_{2, count}^{(h)}y_{p, i}))\\
    f^*(\Theta^{(h)}) =&\phi(y_{cont, i},\beta_{0, cont}^{(h)} + \beta_{1, cont}^{(h)}y_{p, i}, \sigma^{(h)})\\&P(y_{count, i}, link(\beta_{0, count}^{(h)} + \beta_{1, count}^{(h)}y_{count, i} + \beta_{2, count}^{(h)}y_{p, i})).
\end{align*}
These quantities can thus be calculated by the following R code:
\begin{Verbatim}[frame=single]
f_h = dnorm(y_cont_guess, b0_cont[h] + b1_cont[h] * y_p_i, sigma[h]) 
      * dpois(y_count_guess, exp(b0_count[h]
              + b1_count[h] * y_cont_guess + b2_count[h] * y_p_i)))
f_hs = dnorm(y_cont_i, b0_cont[h] + b1_cont[h] * y_p_i, sigma[h])
       * dpois(y_count_i, exp(b0_count[h] + b1_count[h] * y_cont_i 
                              + b2_count[h] * y_p_i)))
\end{Verbatim}
Here \texttt{y\_cont\_guess} and \texttt{y\_count\_guess} are the guesses currently being evaluated for record $i$ and \texttt{y\_cont\_i} and \texttt{y\_count\_i} are its true confidential values. 

With $g(\Theta^{(h)})$, $f(\Theta^{(h)})$, and $f^*(\Theta^{(h)})$ calculated, we now proceed to describe the computational details of the importance sampling step.

\subsection{Implementation details of importance sampling}
\label{methods:importance}
Section \ref{methods:computing} illustrate with sample R script how to calculate $g(\Theta^{(h)})$, $f(\Theta^{(h)})$, and $f^*(\Theta^{(h)})$ for three different Bayesian synthesizers. As reviewed in Section \ref{methods:AR_cal}, once these quantities are calculated, the final step is to use importance sampling to approximate the posterior probability of each guess for record $i$, through Equation (\ref{eq:sampling}). 

We present the implementation details of the importance sampling step with illustration to the two synthetic continuous variables example in Section \ref{methods:computing:2cont}. The implementation details for other synthesizers follow a similar structure and are omitted for brevity.  

The following sample R script approximates the risk for one pair of guesses (\texttt{y\_1\_guess} and \texttt{y\_2\_guess}) for record $i$: 


\begin{Verbatim}[frame=single]
f_k = dnorm(y_1_guess, mu, sigma_1) * 
      dnorm(y_2_guess, b0 + y_1_guess * b1, sigma_2) 
f_ks = dnorm(y_1_i, mu, sigma_1) * 
                dnorm(y_2_i, b0 + y_1_i * b1, sigma_2)
  
denom = sum(f_k / f_ks)

for (h in 1:H) {
  g_h = dnorm(y_1_syn, mu[h], sigma_1[h]) *
        dnorm(y_2_syn, b0[h] + y_1_syn * b1[h], sigma_2[h])
  g_h = prod(g_h)
     
  f_h = dnorm(y_1_guess, mu[h], sigma_1[h]) *
        dnorm(y_2_guess, b0[h] + y_1_guess * b1[h], sigma_2[h])
  f_hs = dnorm(y_1_i, mu[h], sigma_1[h]) *
         dnorm(y_2_i, b0[h] + y_1_i * b1[h], sigma_2[h])
     
  post_prob[h] = g_h * (f_h / f_hs) / denom
}
post_prob = mean(post_prob)
\end{Verbatim}
Here \texttt{y\_1\_i}, \texttt{y\_2\_i}, \texttt{y\_1\_guess}, and \texttt{y\_2\_guess} are scalars for record $i$, while \texttt{y\_1\_syn} and \texttt{y\_2\_syn} are synthetic vectors of length $n$ all $n$ records and \texttt{b0}, \texttt{b1}, \texttt{mu}, \texttt{sigma\_1}, and \texttt{sigma\_2} are vectors of length $H$.  

To compute the attribute disclosure risks for record $i$, we run the above script for $G_1 \times G_2$ guesses in the collection of $\{y_{11}^*, \cdots, y_{1G_1}^*, y_{21}^*, \cdots, y_{2G_2}^*\}$. The final posterior probability of each guess is scaled with the sum of $G_1 \times G_2$ guesses for re-normalization. This process is repeated for every target record. In addition to the posterior probability of the true confidential values, summaries such as how the confidential values rank among all the guesses can be reported as attribute disclosure risks. 

\section{The \texttt{AttributeRiskCalculation} Package}
\label{package}

We have created and made public the \texttt{AttributeRiskCalculation} R package, which calculates attribute disclosure risks for synthetic datasets with the methods discussed in Section \ref{methods}. We describe how to use the package, specifically what the \texttt{AttributeRisk()} function computes, the inputs it requires, and the outputs it produces. The following R code shows how to install the package.


\begin{Verbatim}[frame=single]
library(devtools)
install_github("RyanHornby/AttributeRiskCalculation")
\end{Verbatim}

The \texttt{AttributeRisk()} function in this R package computes the attribute disclosure risks for \emph{all} records in the synthetic dataset. Its outputs include the joint posterior probability matrix, the marginal probabilities each variable (if more than one variable is synthesized), the rank of the true value among the all guesses, and the absolute difference between the guess that has the highest risk, $y^{**}$, and the true confidential value: $|y^{**} - y_i|$. If focusing on one target record, the \texttt{AttributeRiskForRecordI()} function can be used to calculate these quantities. In addition to the regular outputs, it provides the ranks of all the values in the joint posterior probability matrix. 
The dimensions of the joint posterior probability matrix depend on the number of synthesized variables. For example if only one variable is synthesized the output is a vector, while three synthesized variables result in a three dimensional array. 

An example usage of the \texttt{AttributeRisk()} function is below: 
\begin{Verbatim}[frame=single]
AttributeRisk(modelFormulas, 
              origdata, 
              syndata, 
              posteriorMCMCs,
              syntype,
              G,
              H)
\end{Verbatim}

The first required argument \texttt{modelFormulas} is a list of formula or formula like objects (for example \texttt{brmsformula} object in the \texttt{brms} R package \citep{brms}). The elements in this list should appear in the order they were synthesized. For example if $\y_1$ is synthesized with no predictors and then $\y_2$ is synthesized with $\y_1$ as a predictor, \texttt{modelFormulas} should be set to \texttt{list(formula(y1$\sim$1), formula(y2$\sim$y1))}. 

The second required argument \texttt{origdata} is the confidential dataset in a data frame. The third required argument \texttt{syndata} is a list containing synthetic dataset(s). 

The fourth required argument \texttt{posteriorMCMCs} is a list, in order of synthesis, of the MCMC draws from the synthesizer. The last required argument \texttt{syntype} is a vector, in order of synthesis, of strings corresponding to the type of synthesizer used. For example if the first sensitive variable is synthesized with a linear regression synthesizer without predictors and the second is a Poisson regression synthesizer, \texttt{syntype} should be set to \texttt{c("norm", "pois")}. 

Optional arguments mainly allow customization of the guesses chosen by the user, of which the \texttt{AttributeRisk()} function will compute attribute disclosure risks for. For example, the input \texttt{G} is the number of guesses (including the confidential value), and it has a default value of 11 for continuous variables. For categorical or count variables, the function will use all possible guesses for this variable in the dataset. 
The value of $H$, the number of posterior parameter draws for the importance sampling step, can be modified with input \texttt{H} which has a default value of 50. The chosen $H$ should not exceed the number of available MCMC draws of $\Theta$ provided in argument \texttt{posteriorMCMCs}. 

\section{Applications to a CE sample}
\label{app}



We apply our attribute disclosure risks evaluation methods to three applications to a CE sample, each corresponds to an illustrative example in Section \ref{methods:computing}. For synthesis, we use either the \texttt{stan\_glm()} function from the \texttt{rstanarm} R package or the \texttt{brm()} function from the \texttt{brms} R package to fit our synthesizers on the confidential CE data, which provide us with MCMC draws of the model parameters. Code used to synthesize the CE data are included in the Appendix. For illustration purpose, in each application we generate a single synthetic dataset.

Our CE sample comes from the 2019 1st quarter with $n = 5126$ consumer units (CUs). There are 5 variables in this sample with details in Table \ref{tab:CEvars}. 

\begin{table}[H]
    \centering
    \begin{tabular}{p{0.8in} p{0.8in} p{3.5in}}
    \hline Variable & Type &Description \\\hline
    Urban & Categorical & Whether this CU located in an urban or rural area (2 levels). \\
    Race & Categorical & Race category of the reference person (6 levels). \\
    KidsCount & Count & Number of CU members under age 16. \\
    Expenditure & Continuous & Total expenditure last quarter. \\
    Income & Continuous & Total amount of family income before taxes in the last 12 months.\\\hline 
\end{tabular}
\vspace{1mm}
    \caption{Variables used from the CE data sample.}
    \label{tab:CEvars}
\end{table}



When calculating attribute disclosure risks for our applications, we use default $H = 50$ for the number of posterior parameter draws in the importance sampling step. For continuous synthetic variables, we use the default $G = 11$ guesses (the true confidential value plus 10 guesses in the neighborhood within a 20\% range of the true confidential value). This can be done with the following R code: 
\begin{Verbatim}[frame=single]
y_i_guesses  = seq(y_i*0.9, y_i*1.1, length.out = 11) 
\end{Verbatim}

\subsection{Evaluation of synthetic Race}
\label{app:oneCat}


Our first application synthesizes categorical Race with a multinomial logistic regression synthesizer with the LogIncome (the logarithm of Income) as a predictor, see Figure \ref{fig:Race_utility} for the utility evaluation of the resulting synthetic dataset.

\begin{Figure}
    \centering
    \includegraphics[width=0.5\linewidth]{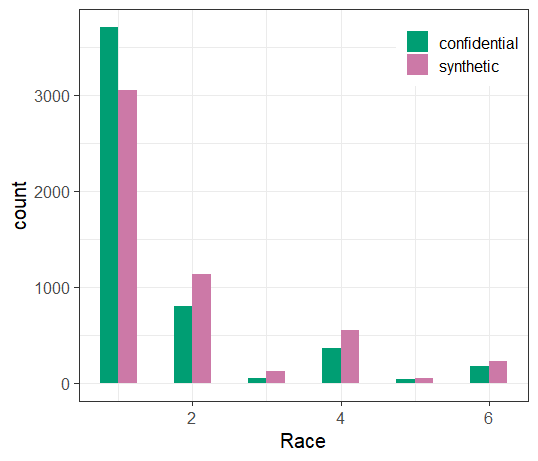}
    \captionof{figure}{Histogram of the confidential and synthetic values of Race.}
    \label{fig:Race_utility}
\end{Figure}

To estimate attribute disclosure risks for each record in the synthetic dataset, we use the \texttt{AttributeRisk()} function the following manner. We provide a list of one formula that describes our synthesis of Race given LogIncome,the confidential dataset (\texttt{CEdata}), the synthetic dataset (\texttt{CEdata\_syn\_cat}), and MCMC draws (\texttt{draws\_cat}). We use \texttt{c("multinom")} for the synthesizer type for categorical Race and the default value of $H$ (\texttt{H = 50}).

\begin{Verbatim}[frame=single]
One_Cat = AttributeRisk(modelFormulas = list(bf(Race ~ LogIncome)),
                        origdata = CEdata, 
                        syndata = CEdata_syn_cat, 
                        posteriorMCMCs = draws_cat,
                        syntype = c("multinom"),
                        H = 50)
\end{Verbatim}

Figure \ref{fig:Race_Prob} shows the density of the re-normalized probabilities of the true confidential values being guessed correctly for all $n = 5126$ records. The results show for majority of the records, its posterior probability is lower than the prior (i.e. randomly guessing among the guesses with probability of 1/6), suggesting low attribute disclosure risks in the synthetic dataset. Figure \ref{fig:Race_rank} shows the rank of the true confidential values among 6 guesses for all records. Most records have rank 2 (rank 1 is highest risk). We have an overall low attribute disclosure risks.


\begin{multicols}{2}
\begin{Figure}
    \centering
    \includegraphics[width=\linewidth]{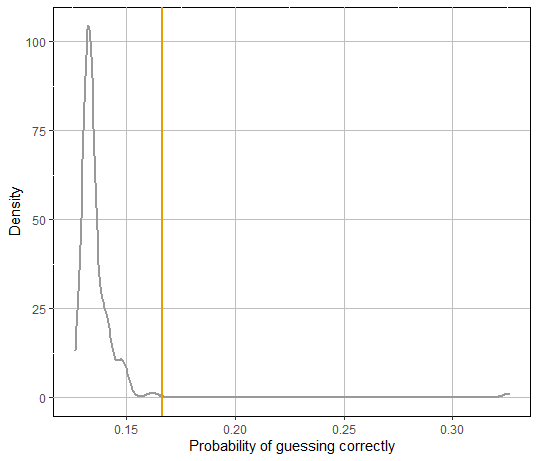}
    \captionof{figure}{Density of the posterior probability of correctly guessing the true confidential value of Race for all CUs. The vertical line shows the prior probability of 1/6.}
    \label{fig:Race_Prob}
\end{Figure}
\begin{Figure}
    \centering
    \includegraphics[width=\linewidth]{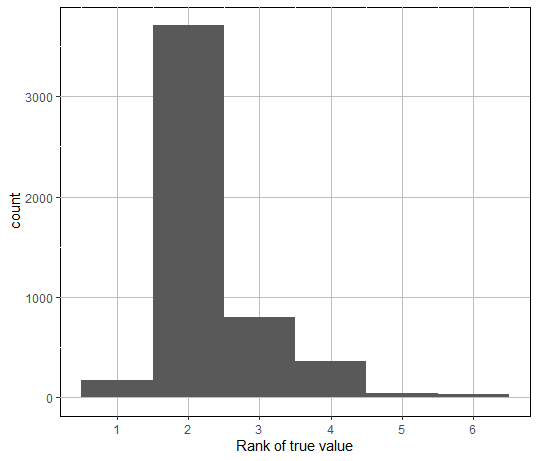}
    \captionof{figure}{The rank of posterior probability of the true value of Race being guessed correctly, among 6 guesses, for all CUs.}
    \label{fig:Race_rank}
\end{Figure}
\end{multicols}

\subsection{Evaluation of synthetic Expenditure and Income}
\label{app:twoCont}

In our second application, we first synthesize LogExpenditure (the logarithm of Expenditure) with a linear regression synthesizer with no predictors. Next, we synthesize LogIncome with another linear regression synthesizer using LogExpenditure as a predictor. Figures \ref{fig:income_expend_util2} and \ref{fig:income_expend_util1} present utility plots of the two synthetic variables, respectively. It is expected that LogExpenditure, being the first variable in the sequential synthesis process, has higher utility, since LogIncome is synthesized given \emph{synthetic} values of LogExpenditure.

\begin{multicols}{2}
\begin{Figure}
    \centering
    \includegraphics[width=\linewidth]{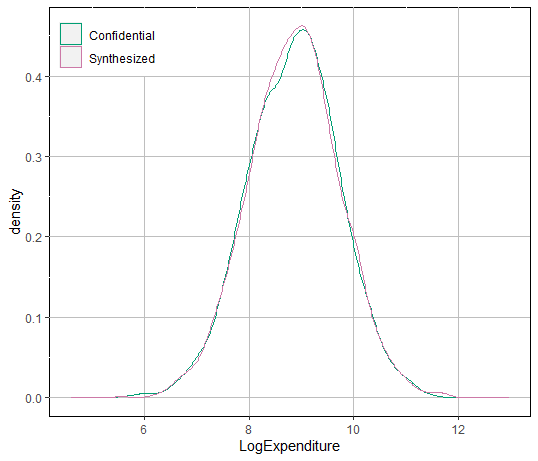}
    \captionof{figure}{Density plot of the confidential and synthetic values of LogExpenditure.}
    \label{fig:income_expend_util2}
\end{Figure}
\begin{Figure}
    \centering
    \includegraphics[width=\linewidth]{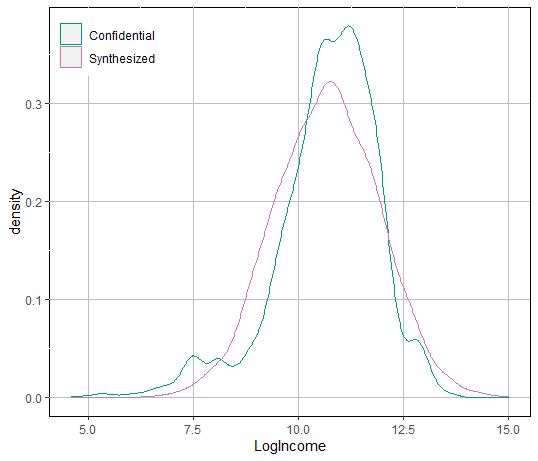}
    \captionof{figure}{Density plot of the confidential and synthetic values of LogIncome.}
    \label{fig:income_expend_util1}
\end{Figure}
\end{multicols}

To estimate attribute disclosure risks for each record in the synthetic dataset, we use the \texttt{AttributeRisk()} function the following manner. We provide a list of the two formulas describing the sequential synthesis process: first LogExpenditure and second LogIncome given LogExpenditure. We also provide the confidential dataset (\texttt{CEdata}), the synthetic data (\texttt{CEdata\_syn\_cont}), and MCMC draws (\texttt{draws\_cont}). Finally we provide the vector of synthesis type, \texttt{c("norm", "norm")}, our choices of $G$ for both continuous variables (\texttt{G = c(11, 11)}), and the value for $H$ (\texttt{H = 50}).

\begin{Verbatim}[frame=single]
Two_Cont = AttributeRisk(modelFormulas = list(bf(LogExpenditure ~ 1),
                                     bf(LogIncome ~ LogExpenditure)),
                         origdata = CEdata, 
                         syndata = CEdata_syn_cont, 
                         posteriorMCMCs = draws_cont, 
                         syntype = c("norm", "norm"), 
                         G = c(11, 11),
                         H = 50)
\end{Verbatim}

Since there are two synthetic variables, we can evaluate the joint posterior probability of guessing the true confidential values of two variables. We can also evaluate their corresponding marginal posterior probabilities, as each marginal probability shows how likely the true confidential value of one variable is being correctly guessed. These marginal probabilities are important to allow us to evaluate how risky it is to correctly guess each individual variable, separately.

\begin{multicols}{2}
\begin{figure}[H]
    \centering
    \includegraphics[width=\linewidth]{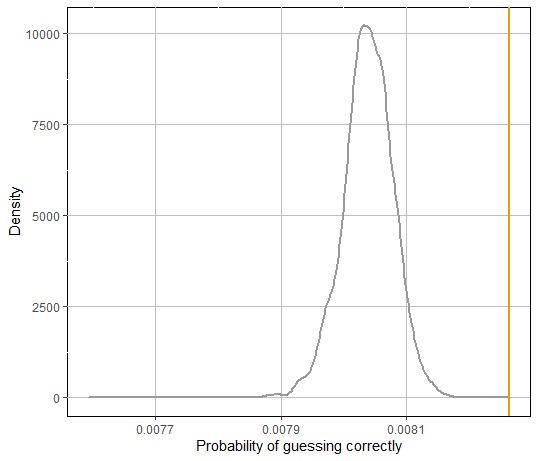}
    \captionof{figure}{Density of the joint posterior probability of correctly guessing the true value of both LogIncome and LogExpenditure. The vertical line shows the prior probability of 1/121.}
    \label{fig:2d_Prob}
\end{figure}

\begin{figure}[H]
    \centering
    \includegraphics[width=\linewidth]{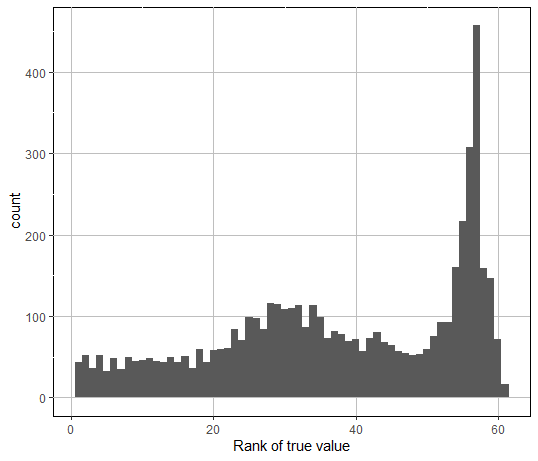}
    \captionof{figure}{The rank of posterior probability of the true pair of values of LogIncome and LogExpenditure being guessed correctly, among 121 guesses, for all CUs.}
    \label{fig:my_fig}
\end{figure}
\end{multicols}

Figure \ref{fig:2d_Prob} is the joint posterior probability. It shows that almost all records have lower joint posterior probability than the prior (the prior is 1/121 since both variables have $G = 11$ guesses). 
Figure \ref{fig:my_fig} shows the rank of the true confidential pair being guessed correctly among 121 guesses for all records (rank = 1 indicates highest attribute disclosure risks). The mode of the rankings is close to 60, showing overall low attribute disclosure risks for the entire dataset. Nevertheless, there are about 50 out of $n = 5126$ records ranked 1st, indicating high attribute disclosure risks for these records.

\begin{figure}[H]
    \centering
    \includegraphics[width=\linewidth]{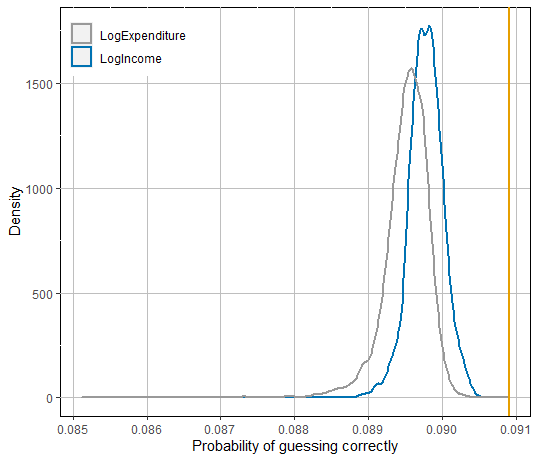}
    \captionof{figure}{Density of the marginal posterior probabilities of correctly guessing the true value of LogIncome and LogExpenditure, respectively. The vertical line shows the prior probability of 1/11.}
    \label{fig:marginal_Prob}
\end{figure}
\begin{figure}[H]
    \centering
    \includegraphics[width=\linewidth]{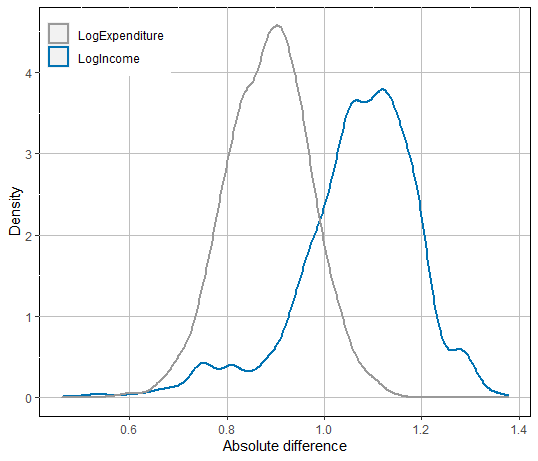}
    \captionof{figure}{Density of the absolute difference between the true value and the guessed value with the largest marginal posterior probability for both LogIncome and LogExpenditure.}
    \label{fig:Income_Abs_Diff_marginal}
\end{figure}

Moving to marginal posterior probabilities in Figure \ref{fig:marginal_Prob}, we can see that LogExpenditure has a slightly lower average chance of being guessed correctly compared to LogIncome. 
Nevertheless, both variables have low attribute disclosure risks (lower than the prior of 1/11). 
Figure \ref{fig:Income_Abs_Diff_marginal} shows the absolute difference between the true confidential value and the highest ranking guess, for LogExpenditure and LogIncome, respectively. On average, the true confidential LogExpenditure is about 0.9 from the highest ranking guess, while the true confidential LogIncome is about 1.1 from the highest ranking guess.  

\begin{multicols}{2}

\end{multicols}

\subsection{Evaluation of synthetic LogExpenditure and KidsCount}
\label{app:1count1bin1cont}

In our final CE application, we first synthesize LogExpenditure with a linear regression synthesizer using a single binary predictor Urban. Next, we synthesize KidsCount with a Poisson regression synthesizer, using two predictors of LogExpenditure and Urban. Figures \ref{fig:kids_expend_util1} and \ref{fig:kids_expend_util2} present the utility plots of the two synthetic variables. As with sequential synthesis in Section \ref{app:twoCont}, the second synthetic variable would have lower utility than the first, exactly what we observe here comparing synthetic KidsCount to synthetic LogExpenditure.

\begin{multicols}{2}
\begin{Figure}
    \centering
    \includegraphics[width=\linewidth]{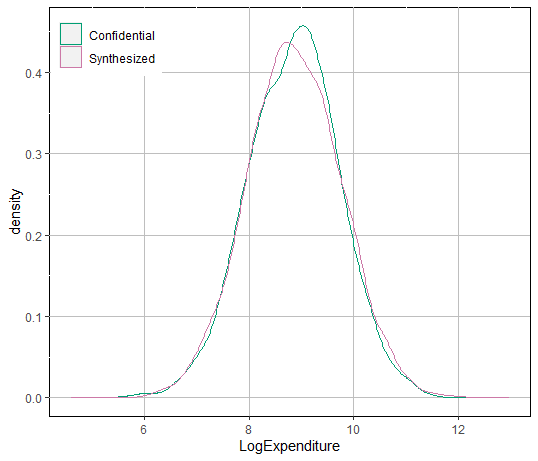}
    \captionof{figure}{Density plot of the confidential and synthetic values of LogExpenditure.}
    \label{fig:kids_expend_util1}
\end{Figure}
\begin{Figure}
    \centering
    \includegraphics[width=\linewidth]{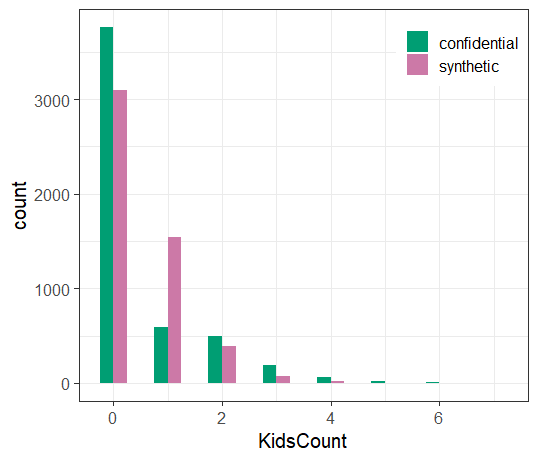}
    \captionof{figure}{Histogram of the confidential and synthetic values of KidsCount.}
    \label{fig:kids_expend_util2}
\end{Figure}
\end{multicols}

To estimate attribute disclosure risks for each record in the synthetic dataset, we use the \texttt{AttributeRisk()} function the following manner. We provide a list of the two formulas describing the synthesis, following the synthesis order. Moreover, we provide the confidential dataset (\texttt{CEdata}), the synthetic data (\texttt{CEdata\_syn\_count}), and MCMC draws (\texttt{draws\_count}). We also provide the vector of synthesis type \texttt{c("norm", "pois")}. Lastly we provide \texttt{G = 11} for the continuous LogExpenditure and \texttt{H = 50} for $H$. 

\begin{Verbatim}[frame=single]
Count_risks = AttributeRisk(
                modelFormulas = list(bf(LogExpenditure ~ Urban),
                     bf(KidsCount ~ LogExpenditure + Urban)),
                origdata = CEdata, 
                syndata = CEdata_syn_count, 
                posteriorMCMCs = draws_count,
                syntype = c("norm", "pois"), 
                G = 11,
                H = 50)
\end{Verbatim}



\begin{figure}[H]
    \centering
    \includegraphics[width=\linewidth]{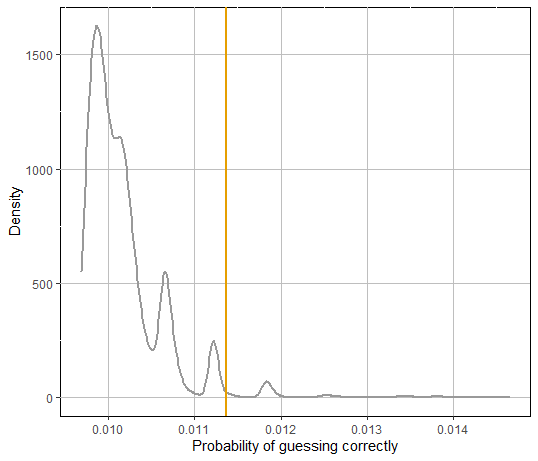}
    \captionof{figure}{Density of the joint posterior probability of correctly guessing the true confidential pair of LogExpenditure and KidsCount. The vertical line shows the prior probability of 1/88.}
    \label{fig:kids_guess}
\end{figure}

\begin{figure}[H]
    \centering
    \includegraphics[width=\linewidth]{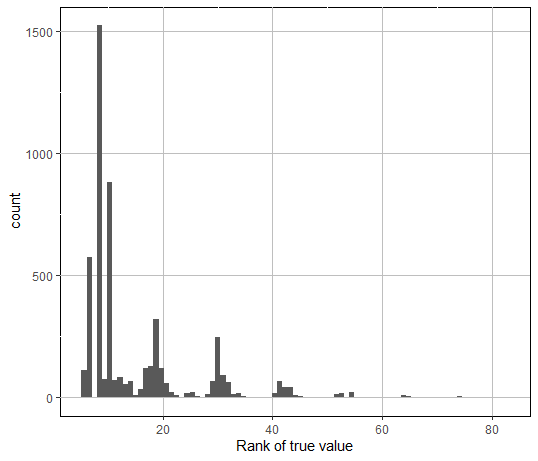}
    \captionof{figure}{The rank of posterior probability of the true pair of values of LogExpenditure and KidsCount being guessed correctly, among 88 guesses, for all CUs. }
    \label{fig:kids_hist}
\end{figure}

\begin{figure}[H]
    \centering
    \includegraphics[width=\linewidth]{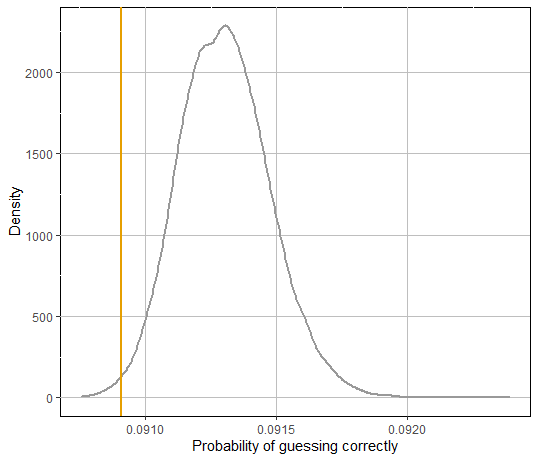}
    \captionof{figure}{Density of the probability of correctly guessing the true value LogExpenditure. The vertical line shows prior probability of 1/11.}
    \label{fig:kids_marginal_expend}
\end{figure}

\begin{figure}[H]
    \centering
    \includegraphics[width=\linewidth]{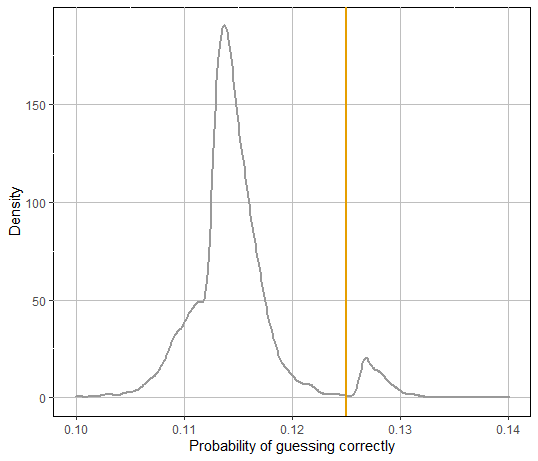}
    \captionof{figure}{Density of the probability of correctly guessing the true value KidsCount. The vertical line shows prior probability of 1/8.}
    \label{fig:kids_marginal_kids}
\end{figure}

Figure \ref{fig:kids_guess} shows the density of the joint posterior probabilities for correctly guessing the true confidential pair. For most records, its posterior probability is lower than the prior of 1/88. The rank plot in Figure \ref{fig:kids_hist} shows a mode around 5. Looking at the marginal probabilities in Figures \ref{fig:kids_marginal_expend} and \ref{fig:kids_marginal_kids}, we can see that guessing the true value of LogExpenditure is actually more likely on average than randomly guessing. This may motivate data disseminators to change their synthesis models if this risk is deemed too large.

\begin{multicols}{2}
\end{multicols}

\begin{multicols}{2}
\end{multicols}


\begin{multicols}{2}
\begin{figure}[H]
    \centering
    \includegraphics[width=\linewidth]{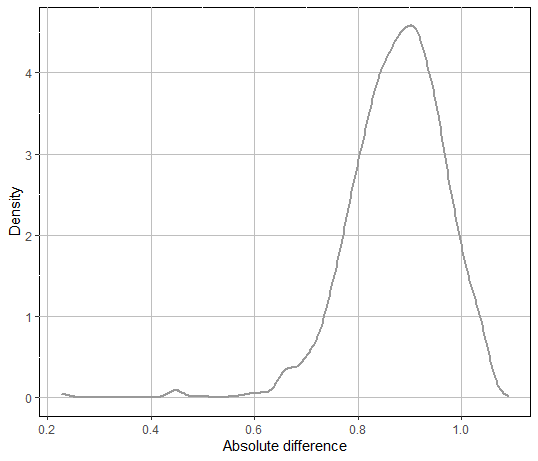}
    \captionof{figure}{Density of the absolute difference between the true value and the guessed value with the largest marginal posterior probability for LogExpenditure.}
    \label{fig:kids_diff_expend}
\end{figure}
\begin{figure}[H]
    \centering
    \includegraphics[width=\linewidth]{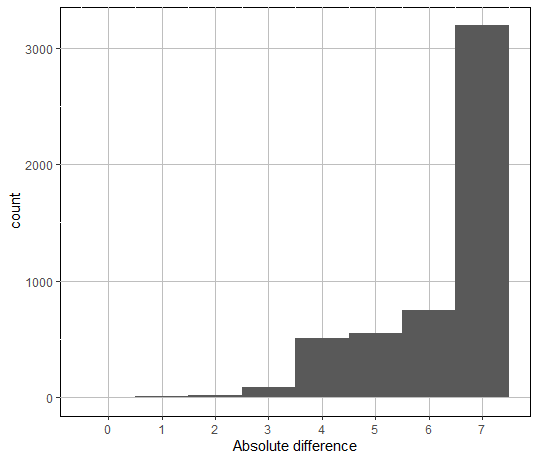}
    \captionof{figure}{Histogram of the absolute difference between the true value and the guessed value with the largest marginal posterior probability for KidsCount.}
    \label{fig:kids_diff_kids}
\end{figure}
\end{multicols}

Lastly for the absolute difference between the true confidential value and the highest ranking guess in Figure \ref{fig:kids_diff_expend} for LogExpenditure and Figure \ref{fig:kids_diff_kids} for KidsCount, LogExpenditure on average has a distance of 0.9, while KidsCount shows more than 3000 CUs are 7 count away from the highest ranking guess (the range for KidsCount is 8), indicating low risks.

\section{Concluding remarks}
\label{conclusion}

In this paper, we reviewed a general approach for calculating attribute disclosure risks using Bayesian estimation methods. For accessible presentation, we provided examples of several commonly-used Bayesian synthesizers with sample R scripts to illustrate the computational details of the estimation methods. We introduced the \texttt{AttributeRiskCalculation} R package for calculating calculate attribute disclosure risks in synthetic datasets, with several use cases to CE sample applications.

The estimation method we outlined in this work requires posterior draws of model parameters in the chosen Bayesian synthesizer(s). Common Bayesian MCMC estimation software usually provide posterior parameter draws in their output like \texttt{stan} or \texttt{JAGS} \citep{rstanarm, Plummer03jags:a}. However, other data synthesizers, such as classification and regression trees (CART), do not involve posterior parameter draws of synthesis models \citep{synthpopJSS}. How to evaluate attribute disclosure risks for these synthesizers is in important future research direction.

Another future research direction is the relaxation of our worst case scenario assumption of a very knowledgeable intruder. Such assumption could be too conservative in practice, although it is necessary for the purpose of efficient computation. Advancement of computation efficiency could afford more realistic assumptions of intruder's knowledge and behavior.



\bibliography{bibliography}
\bibliographystyle{apalike}

\newpage
\section*{Appendix}



We provide our R script used in Section \ref{app} for synthesizing the CE dataset. 

\subsection*{1. Synthesizing categorical Race with the \texttt{brm()} function}

\begin{Verbatim}[frame=single]
syn_multinomial_brms = function(orig_data, syn_data, 
    model_brms = bf(outcome ~ 1), chains = 1, iterations = 1000,
    c = 0.95, thresh = 1.00, m = 20, thin = 5) {
  
  ff = stats::as.formula(model_brms)
  model = stats::model.frame(ff, orig_data)
  X = data.frame(stats::model.matrix(ff, model))
  
  multi_logistic_fit = brms::brm(data = orig_data,
                                 family = categorical(link="logit"),
                                 model_brms,
                                 iter = iterations,
                                 chains = chains)
  
  post_multi_logistic = brms::posterior_samples(multi_logistic_fit)
  
  n = length(orig_data[,1])
  C = length(levels(orig_data[, 
                        paste(text = model_brms$formula[[2]])]))
  start = length(post_multi_logistic[,1]) - thin  * (C - 1)
  log_p_allC = matrix(NA, nrow = n, ncol = C)
  for (c in 2:C){
    name_Intercept_c = paste0("b_mu", c, "_Intercept")
    name_LogIncome_c = paste0("b_mu", c, "_LogIncome")
    index = start + thin * (c - 2)
    log_p_c = as.matrix(X) %*% 
      t(post_multi_logistic[index, 
                        c(name_Intercept_c, name_LogIncome_c)])
    log_p_allC[, c] = log_p_c
  }
  log_p_allC[, 1] = rep(0, n)
  
  p_allC = exp(log_p_allC) / (1 + exp(log_p_allC))
  
  syndata = vector("list", m)
  for (i in 1:m){
    synthetic_Y = rep(NA, n)
    for (i in 1:n){
      synthetic_Y[i] = which(rmultinom(1, size = 1, 
                                    prob = p_allC[i, ]) == 1)
    }
    syndata[[i]] = synthetic_Y
  }
  
  return(list(syndata, p_allC))
}

CEdata_syn_cat = CEdata
CEdata$Race = as.factor(CEdata$Race)

draws_cat = list()
synthesis_cat = syn_multinomial_brms(CEdata, CEdata_syn_cat, 
                                     bf(Race ~ LogIncome), m = 1)

CEdata_syn_cat$Race = synthesis_race[[1]][[1]]
draws_cat[[1]] = synthesis_race[[2]]

CEdata_syn_cat = list(CEdata_syn_cat)
\end{Verbatim}

\subsection*{2. Synthesizing continuous LogExpenditure and LogIncome sequentially with the \texttt{stan\_glm()} function}

\begin{Verbatim}[frame=single]
syn_normal_brms = function(orig_data, syn_data, 
                   model_brms = brmsformula(outcome ~ 1), 
                   chains = 1, iterations = 1000, m = 20, thin = 5) {
  ff = as.formula(model_brms)
  utils::str(model <- model.frame(ff, syn_data))
  X = model.matrix(ff, model)
  
  fit = stan_glm(
    model_brms,
    data = orig_data,
    family = gaussian(),
    prior = normal(0, 2, autoscale = FALSE),
    refresh = 0,
    chains = chains, iter = iterations
  )
  
  #### synthesis ####
  N = length(orig_data[,1])
  draws = as.data.frame(fit)
  start = length(draws[,1]) - thin  * (m - 1)
  syndata = vector("list", m)
  for (i in 1:m){
    indx = start + thin * (i - 1)
    
    draws_exp_mean = as.matrix(X) %*% 
                      t(draws[indx, !names(draws) %in% c("sigma")])
    
    draws_sd = draws[indx, "sigma"]
    syndata[[i]] = rnorm(N, mean = draws_exp_mean, sd = draws_sd)
  }
  return(list(syndata, draws))
}
CEdata_syn_cont = CEdata
draws_cont = list()

synthesis_cont = syn_normal_brms(CEdata, CEdata_syn_cont, 
                            bf(LogExpenditure ~ 1), m = 1)
CEdata_syn_cont$LogExpenditure = synthesis_cont[[1]][[1]]
synthesis_cont2 = syn_normal_brms(CEdata, CEdata_syn_cont, 
                            bf(LogIncome ~ LogExpenditure), m = 1)

CEdata_syn_cont$LogIncome = synthesis_cont2[[1]][[1]]
draws_cont[[1]] = synthesis_cont[[2]]
draws_cont[[2]] = synthesis_cont2[[2]]

CEdata_syn_cont = list(CEdata_syn_cont)
\end{Verbatim}

\subsection*{3. Synthesizing continuous LogExpenditure and count KidsCount sequentially with the \texttt{stan\_glm()} function}

\begin{Verbatim}[frame=single]
syn_pois_brms = function(orig_data, syn_data, 
        model_brms = bf(outcome ~ 1), chains = 1, iterations = 1000,
        c = 0.95, thresh = 1.00, m = 20, thin = 5) {
  ff = as.formula(model_brms)
  utils::str(model <- model.frame(ff, syn_data))
  X = model.matrix(ff, model)
  
  fit = stan_glm(
    model_brms,
    data = orig_data,
    family = poisson(link = "log"),
    prior = normal(0, 2, autoscale = FALSE),
    refresh = 0,
    chains = chains, iter = iterations
  )
  
  #### synthesis ####
  N = length(orig_data[,1])
  draws = as.data.frame(fit)
  start_draws = length(draws[,1]) - thin  * (m - 1)
  start_data = length(syn_data[,1]) - thin  * (m - 1)
  syndata = vector("list", m)
  
  for (i in 1:m){
    indx = start_draws + thin * (i - 1)
    
    draws_exp_mean = exp(as.matrix(X) %*% t(draws[indx, ]))
    
    syndata[[i]] = rpois(N, lambda = draws_exp_mean)
  }
  
  return(list(syndata, draws))
}
CEdata_syn_count = CEdata
draws_count = list()

synthesis_count = syn_normal_brms(CEdata, CEdata_syn_count, 
    bf(LogExpenditure ~ as.factor(UrbanRural)), m = 1)
CEdata_syn_count$LogExpenditure = synthesis_count[[1]][[1]]
synthesis_count2 = syn_pois_brms(CEdata, CEdata_syn_count, 
    bf(KidsCount ~ LogExpenditure + as.factor(UrbanRural)), m = 1)

CEdata_syn_count$KidsCount = synthesis_count2[[1]][[1]]
draws_count[[1]] = synthesis_count[[2]]
draws_count[[2]] = synthesis_count2[[2]]

CEdata_syn_count = list(CEdata_syn_count)
\end{Verbatim}



\end{document}